\documentclass[%
reprint,
 amsmath,amssymb,
 aps,
 prd,
 10pt,
]{revtex4-2}

\usepackage{graphicx}
\usepackage{dcolumn}
\usepackage{bm}
\usepackage{hyperref}

\usepackage{subfig}
\usepackage{ragged2e}
\usepackage[compat=1.1.0]{tikz-feynman}
\usepackage{color}
\definecolor{comment}{rgb}{.4,.4,.4}
\usepackage{algorithm}
\usepackage[
noEnd=false,
indLines=false,
italicComments=false,
rightComments=false,
commentColor=comment,
beginComment={~~//~},
endComment={},
]{algpseudocodex}
\algrenewcommand\textproc{\textrm}
\newcommand\hrulex[1]{\rule{\columnwidth}{#1}}
\newcommand\urulex[1]{\vspace{+.5\baselineskip}\hrulex{#1}\vspace{-.5\baselineskip}}
\newcommand\drulex[1]{\vspace{-.5\baselineskip}\hrulex{#1}}

\begin{document}

\preprint{APS/123-QED}

\title{Probing charged lepton flavor violation in an economical muon on-target experiment}%

\author{Leyun \surname{Gao}}%
\email{seeson@pku.edu.cn}%
\author{Zijian \surname{Wang}}%
\author{Cheng-en \surname{Liu}}%
\author{Jinning \surname{Li}}%
\author{Alim \surname{Ruzi}}%
\author{Qite \surname{Li}}%
\author{Chen \surname{Zhou}}%
\author{Qiang \surname{Li}}%
\email{qliphy0@pku.edu.cn}%
\affiliation{School of Physics and State Key Laboratory of Nuclear Physics and Technology, Peking University, Beijing, 100871, China}%

\date{\today}

\begin{abstract}
This work proposes a new yet economical experiment to probe the charged lepton flavor violation (CLFV) process mediated by an extra massive neutron gauge boson $Z^\prime$ beyond the standard model, by extending a recently proposed muon dark matter project in the Peking University Muon (PKMuon) Experiment. The devices used originally for light mass dark matter direct detection are easily adaptable to search for the $\mu^+e^- \to \mu^+\mu^-$ CLFV process leveraging the large-area, high-precision muon tracking and tomography system sandwiching a fixed target the incoming muons scatter off. The $\mu^+\mu^-$ final state signal studied in this work can be uniquely sensitive to specific CLFV parameter combinations, such as the couplings between $Z^\prime$, electron and muon, or $Z^\prime$ and two muons. Prospected results are obtained through detailed detector simulation for the proposal interfacing with a muon beam with energy at tens of $\mathrm{GeV}$ and a flux of $10^6\ \mathrm{s^{-1}}$. Based mainly on angular information of the incoming and outgoing particles, the expected upper limit at 95\% confidence level on the coupling coefficients $\lambda_{e\mu}\lambda_{\mu\mu}$ is able to reach $10^{-5}$ with, for example, $Z^\prime$ mass $0.25\ \mathrm{GeV}$, for a one year's run.

\end{abstract}

\maketitle


\section{Introduction}

In the standard model of particle physics, the charged lepton flavor violation (CLFV) processes are strongly suppressed due to the tiny neutrino masses, hence unobservable yet in current experiments. However, they can be much enhanced in various models beyond the standard model, one of which introduces an additional $U(1)$ gauge symmetry corresponding to a massive neutral gauge boson $Z^\prime$. In this study, similar theoretical claims are assumed as in Ref.~\cite{Ding:2024zaj,Langacker:2000ju} that the extra $Z^\prime$ current has the same gauge coupling and chiral strength as the standard model $Z$ boson except for allowing the charged lepton flavor violations quantitated by
\begin{equation}
\lambda =\left(\begin{array}{lll}
\lambda_{e e} & \lambda_{e \mu} & \lambda_{e \tau} \\
\lambda_{\mu e} & \lambda_{\mu \mu} & \lambda_{\mu \tau} \\
\lambda_{\tau e} & \lambda_{\tau \mu} & \lambda_{\tau \tau}
\end{array}\right).
\label{eq:lambda}
\end{equation}

The $\lambda$ matrix represents the strengths of the CLFV couplings relative to the standard model, with the diagonal elements equaling 1 and the off-diagonal near 0 in expectation. Dedicated low-energy muon experiments, such as $\mu \to e\gamma$, set limits on $\lambda_{ee}\lambda_{e\mu}$ and $\lambda_{e\mu}\lambda_{\mu\mu}$ coherently according to Ref.~\cite{Langacker:2000ju} as
\begin{equation}\label{eq:megwidth}
\begin{split}
\Gamma(\mu \to e\gamma) = \frac
{\alpha G_F^2m_\mu^3M_Z^4\left(\sin^2\theta_W\left(\sin^2\theta_W - 1/2\right)\right)^2}
{4\pi^4M_{Z^\prime}^4}
\\ \left(
\lambda_{ee}\lambda_{e\mu}m_e
+ \lambda_{e\mu}\lambda_{\mu\mu}m_\mu
\,{\color{gray}+\,\lambda_{e\tau}\lambda_{\tau\mu}m_{\tau} + \dots}
\right)^2,
\end{split}
\end{equation}
where $\alpha$ is the fine structure constant, $G_F$ is the Fermi constant, $\theta_W$ is the weak mixing angle, and $M_{Z}$, $M_{Z^\prime}$, $m_e$, $m_\mu$, and $m_\tau$ are the masses of the particles referred to. The gray part of Eq. \eqref{eq:megwidth} contributed by the $\tau$ lepton in the standard model and new physics beyond the standard model is not considered in previous works. However, strong interference and cancellation between the terms shown or omitted in Eq. \eqref{eq:megwidth} are possible, allowing the existence of terms with very large modulus, thus the necessity to exclusively probe each single term therein.

\begin{figure}[t]
\includegraphics[width=.8\columnwidth]{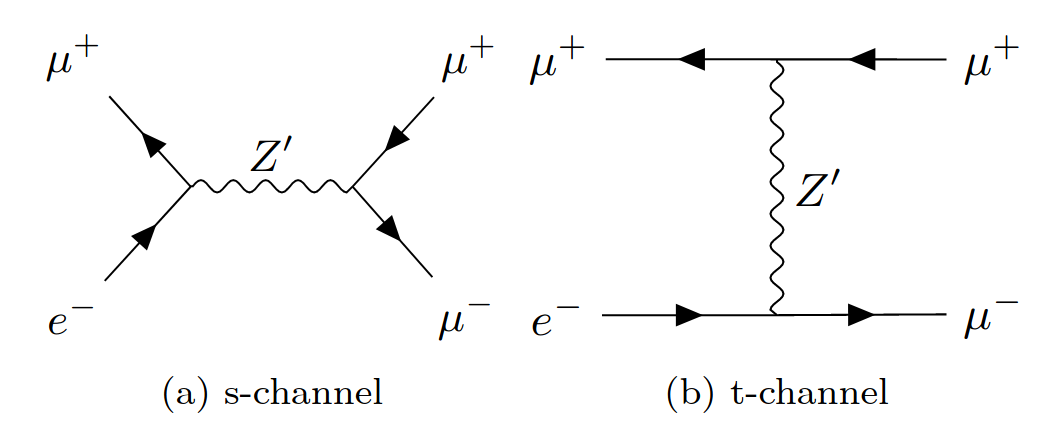}
\caption{\justifying $\mu^+e^- \to \ell^+\ell^-$ production diagrams, taking $\ell = \mu$ for example.}
\label{fig:muemumu-feynman}
\end{figure}

In this paper, the $Z^\prime$-mediated CLFV process $\mu^+e^- \to \ell^+\ell^-$ shown in Fig.~\ref{fig:muemumu-feynman} is probed in simulation with muon beams hitting electrons in a fixed target tracked by sandwiching detectors. The Peking University Muon Experiment that this work belongs to and where the devices and technicals are inherited is introduced in Sec. \ref{sec:pkmuon}. An efficient CLFV Monte Carlo event generation algorithm tailored for the sandwiching detector simulation is discussed in Sec. \ref{sec:signal}. Muon-material interaction simulation, including the background modeling and the detector construction, is described in Sec. \ref{sec:simulation}. Detector-specified lepton track reconstruction and signal purification procedures are discussed in Sec. \ref{sec:selection}. Results are shown in Sec. \ref{sec:results} including the expected 95\% confidence level upper limit on the coupling coefficients $\lambda_{e\mu}\lambda_{\mu\mu}$ for a one year's run, followed by the summary and outlook presented in Sec. \ref{sec:summary}. Details about the kinematics of outgoing $\ell^+\ell^-$ are given in Appendix \ref{sec:kinematic}.

\section{The Peking University Muon Experiment}\label{sec:pkmuon}

Ref.~\cite{Yu:2024spj} proposed a new device to directly detect light mass dark matter through its scattering with abundant atmospheric muons or accelerator beams. A first plan is to use the free cosmic-ray muons interacting with dark matter in a volume surrounded by tracking detectors, to trace possible interaction between dark matter and muons.

\begin{figure}[t]
\includegraphics[width=.25\columnwidth]{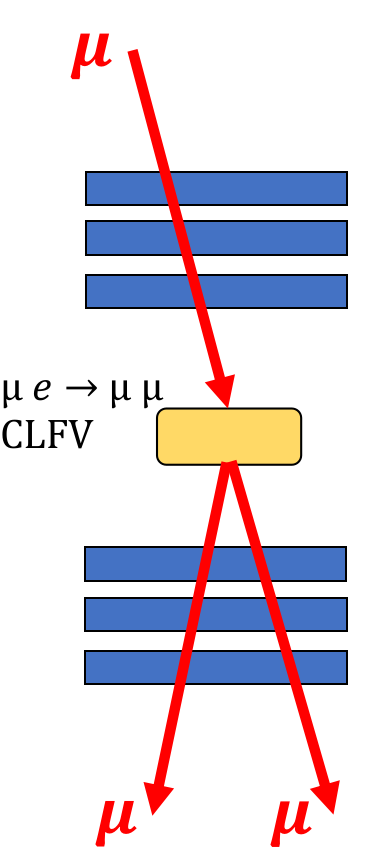}
\caption{\justifying A proposed muon experiment to probe CLFV through muon on-target collisions, which shows only minimal sets of detectors to record muon hits and tracks, with flexibility to include more detector components to cover more physics cases in the future. }
\label{fig:pkmuon-clfv}
\end{figure}

Such a proposal can be extended to cover many other physics cases, such as the CLFV processes. Fig.~\ref{fig:pkmuon-clfv} shows the carton for a proposed muon experiment to probe CLFV through muon on-target collisions. The blue layers represent the muon detectors, such as Resistive Plate Chambers (RPCs). The yellow block is for the target with electrons inside being collided by cosmic muons or muon beams. A particle identification (PID) system made by scintillators below the RPCs for the outgoing muon tracks is a possible future extension to suppress the electron backgrounds in the CLFV experiment.

\section{Efficient Signal Event Generation}\label{sec:signal}

\begin{figure}[t]
\centering
\subfloat[lab frame]{\label{fig:lab-frame}\includegraphics[width=.5\columnwidth]{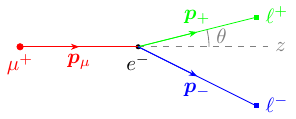}}
\subfloat[COM frame]{\label{fig:com-frame}\includegraphics[width=.5\columnwidth]{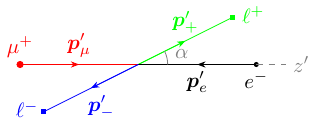}}
\caption{\justifying The $\mu^+e^- \to \ell^+\ell^-$ process in the (a) lab and (b) COM frames.}
\label{fig:frames}
\end{figure}

MadGraph5\_aMC@NLO 3.5.5~\cite{Alwall:2014hca} is exploited to simulate the $\mu^+e^- \to \ell^+\ell^-$ process induced by the $Z^\prime$ boson, where the target electron is set as static in the lab frame and the muon energy $E_\mu$ and the $Z^\prime$ mass $m_{Z^\prime}$ are varied. The off-diagonal in Eq. \eqref{eq:lambda} are set as 1 and can be changed individually in subsequent uses by scaling the Monte Carlo (MC) event weights accordingly. The $Z^\prime$ decay width $\Gamma_{Z^\prime}$ is altered according to $m_{Z^\prime}$, computed with MadGraph5\_aMC@NLO giving $\Gamma_{Z^\prime} \approx 0.02\,m_{Z^\prime}$ for $m_{Z^\prime} < 1\ \mathrm{GeV}$ where the off-diagonal contribute $< 8\%$. All selections on the final state kinematics are disabled. For detector simulation described in the next section, generating one event for each encountered $E_\mu$ randomly determined by the muon interactions with the first half of the detector can be prohibitively expensive, so the kinematic rules for a limited number of $E_\mu$-determined grid points will first be acquired by massive pre-generated events and then used for interpolation inside the whole parameter space. Although the idea applies to both the $e^+e^-$ and $\mu^+\mu^-$ final states and is generalizable to many other physics processes with a low-dimension phase space, the examples below will focus on the $\mu^+e^- \to \mu^+\mu^-$ process of our special interest for the CLFV theory test.

\begin{figure}[t]
\includegraphics[width=.6\columnwidth]{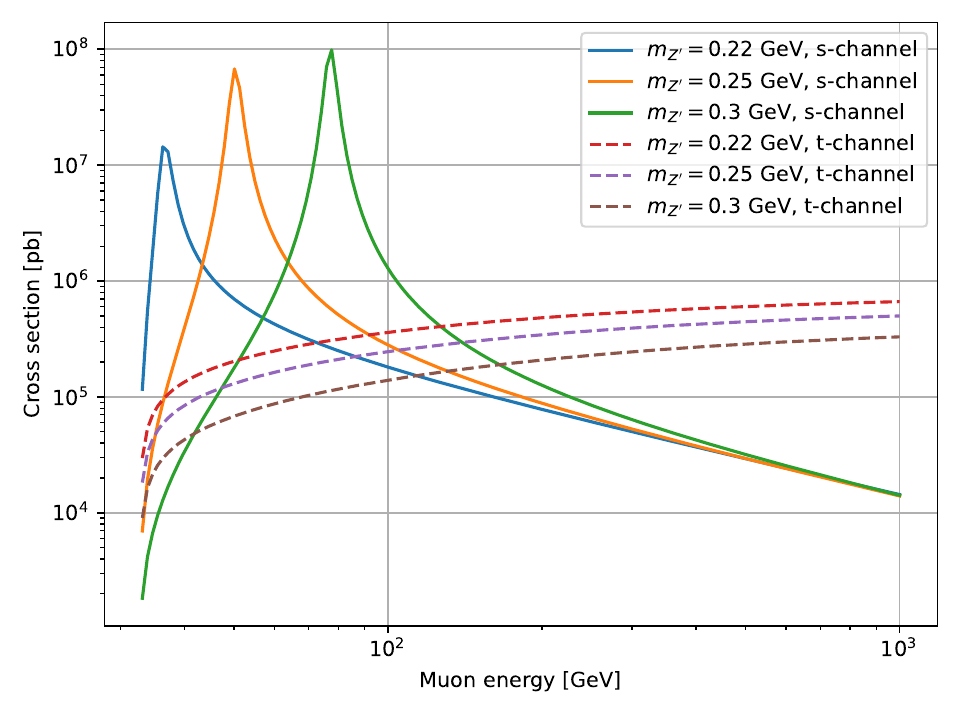}
\caption{\justifying $\mu^+e^- \to \mu^+\mu^-$ production cross section in variant $(m_{Z^\prime},\ E_\mu)$ configuration, for both Z' mediated s-channel and the non-resonant t-channel cases, respectively.}
\label{fig:muemumu-xs}
\end{figure}

\begin{figure}[t]
\subfloat[$m_{Z^\prime} = 0.22\ \mathrm{GeV}$]{\includegraphics[width=.5\columnwidth]{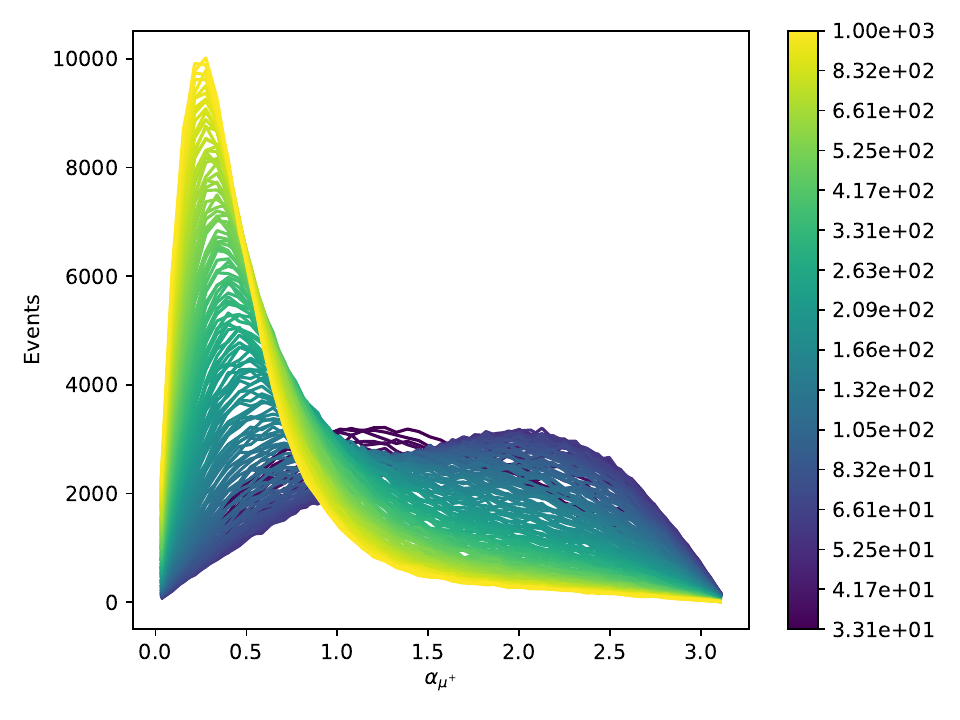}}
\subfloat[$m_{Z^\prime} = 0.3\ \mathrm{GeV}$]{\includegraphics[width=.5\columnwidth]{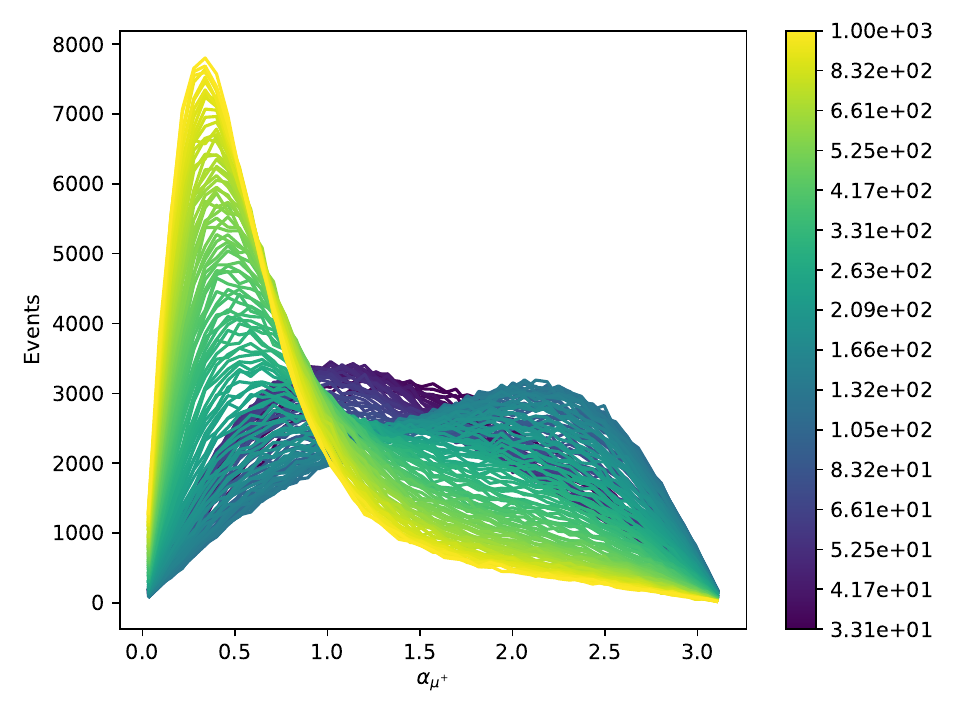}}
\caption{\justifying Distributions of the polar angle of the outgoing $\ell^+$ in the center of mass frame, for $\mu^+e^- \to \mu^+\mu^-$ events with varied $m_{Z^\prime}$, colored by the varying $E_\mu$.}
\label{fig:muemumu-alpha}
\end{figure}

Cross section in different $(m_{Z^\prime},\ E_\mu)$ configuration is plotted in Fig. \ref{fig:muemumu-xs}. The contributions are divided into s- and t-channel as diagrammed in Fig. \ref{fig:muemumu-feynman}. The former dominates in resonance areas where the mediating $Z^\prime$ boson is on-shell. For the reason explained in Appendix \ref{sec:kinematic}, $\alpha$, the polar angle of the outgoing $\ell^+$ in the center of mass (COM) frame as shown in Fig. \ref{fig:com-frame}, is selected exclusively to describe the event distribution in the final state phase space. The $\alpha$ distribution is acquired from the simulated events by constructing a linear histogram ranging $[0, \pi]$. With a given $m_{Z^\prime}$, to illustrate the feasibility of interpolation for arbitrary $E_\mu$ from the grid points, the $\alpha$ distribution histograms of the grid points for typical $m_{Z^\prime}$'s are given in Fig. \ref{fig:muemumu-alpha}. It shows that the $\alpha$ distribution transits contiguously with $E_\mu$ goes from its minimal possible value for the $\mu^+\mu^-$ final state to $1\ \mathrm{TeV}$, allowing acquiring the $\alpha$ distribution for arbitrary $E_\mu$ with linear interpolation between two adjacent grid points.

With the cross section and $\alpha$ distribution grid points, given an $m_{Z^\prime}$ of interest, $\mu^+e^- \to \ell^+\ell^-$ events with arbitrary incoming muon energy-momentum $(E_\mu, \vec p_\mu)$ can be efficiently generated with Algorithm \ref{alg:interp}, where the constants $\sigma_i$ and $H_i$ are respectively the cross section and $\alpha$ distribution histogram corresponding to the energy point $E_i\ (i = 1, 2)$. Details about the variables describing the kinematics in the COM frame $p^\prime$ and $E^\prime$ and the ones describing the Lorentz transformation between the lab and COM frame $\gamma$ and $\beta$ can be referred to in Appendix \ref{sec:kinematic}. The ROOT framework \cite{Brun:1997pa} is exploited for histogram making, storing, and sampling in the algorithm implementation.

\begin{figure}[t]
\urulex{1pt}
\captionof{algorithm}{Efficient $\mu^+e^- \to \ell^+\ell^-$ event generation}
\label{alg:interp}
\drulex{1pt}
\begin{algorithmic}
\Function{GenerateMuELL}{$E_\mu, \vec p_\mu, m_\ell$}
\State $E_{\mu1}, E_{\mu2}, \sigma_1, \sigma_2, H_1, H_2 \gets \Call{AdjacentGridPoints}{E_\mu}$;
\If{$E_\mu$ is out of the grid range}
\State \Return no $\mu^+e^- \to \ell^+\ell^-$ happens;
\EndIf
\State
$w_1 \gets \frac{E_{\mu} - E_{\mu2}}{E_{\mu1} - E_{\mu2}}$,
$w_2 \gets \frac{E_{\mu1} - E_{\mu}}{E_{\mu1} - E_{\mu2}}$;
\State $\sigma \gets w_1\sigma_1 + w_2\sigma_2$;\Comment{cross section}
\If{$\Call{Random}{0, 1} < w_1$}
\State $\alpha \gets H_1.\Call{Sample}{\null}$;
\Else
\State $\alpha \gets H_2.\Call{Sample}{\null}$;
\EndIf\Comment{equivalent to $\alpha \gets (w_1H_1 + w_2H_2).\Call{Sample}{\null}$}
\State $\phi \gets \Call{Random}{0, 2\pi}$;
\State $E^\prime, p^\prime, \gamma, \beta \gets \Call{Kinematics}{E_\mu, m_\ell}$;\Comment{see Appendix \ref{sec:kinematic}}
\State
$p_{x+} \gets p^\prime\sin\alpha\cos\phi$,
$p_{y+} \gets p^\prime\sin\alpha\sin\phi$;
\State $p_{z+} \gets \gamma\left(p^\prime\cos\alpha + \beta E^\prime/2\right)$;
\State $\vec p_+ \gets \Call{ThreeVector}{p_{x+}, p_{y+}, p_{z+}}$;\Comment{see Fig. \ref{fig:lab-frame}}
\State $\vec p_+ \gets \vec p_+.\Call{RotateZAxisTo}{\hat p_\mu}$;
\State $\vec p_- \gets \vec p_\mu - \vec p_+$;\Comment{$\hat z \gets \hat p_\mu$ in Fig. \ref{fig:lab-frame} in rotation}
\State \Return $\sigma, \vec p_+, \vec p_-$;
\EndFunction
\end{algorithmic}
\drulex{1pt}
\end{figure}

\section{Simulation Setup}\label{sec:simulation}

Interactions between the incoming muons with the material in the detector are simulated by GEANT4 11.2.2 \cite{Allison:2016lfl,Allison:2006ve,GEANT4:2002zbu}, where the $\mu^+e^- \to \ell^+\ell^-$ CLFV process is implemented as a discrete process \cite{GEANT4:2023toolkit} upon Algorithm \ref{alg:interp}. An interaction length computing routine is provided to limit the maximum step length allowed for the CLFV process \cite{GEANT4:2023toolkit}. The probability of no CLFV interaction for a $\mu^+$ travelling step length $l$ in the material with mean free path $L = 1 / (n_e\sigma)$ of the process is computed as $\mathrm e^{-l/L}$, where $n_e$ is the material electron density and $\sigma$ is the process cross section. An event is considered a signal (background) event if the CLFV process happens (does not happen) throughout the whole flight. For each $\ell$ and $m_{Z^\prime}$, a global cross section scale factor is applied to give a signal proportion between $10^{-3}$ to $10^{-2}$. It leads to satisfactory simulation efficiency while keeping the events with more than one CLFV process negligible. The interaction length is computed as $10^{-3}$ times the scaled mean free path. It guarantees the simulation granularity without affecting the performance. The signal process is appended to the inclusive physics list FTFP\_BERT \cite{GEANT4:2023physics} provided by GEANT4 where various kinds of hadronic and electromagnetic backgrounds are modeled.

\begin{figure}[t]
\includegraphics[width=\columnwidth]{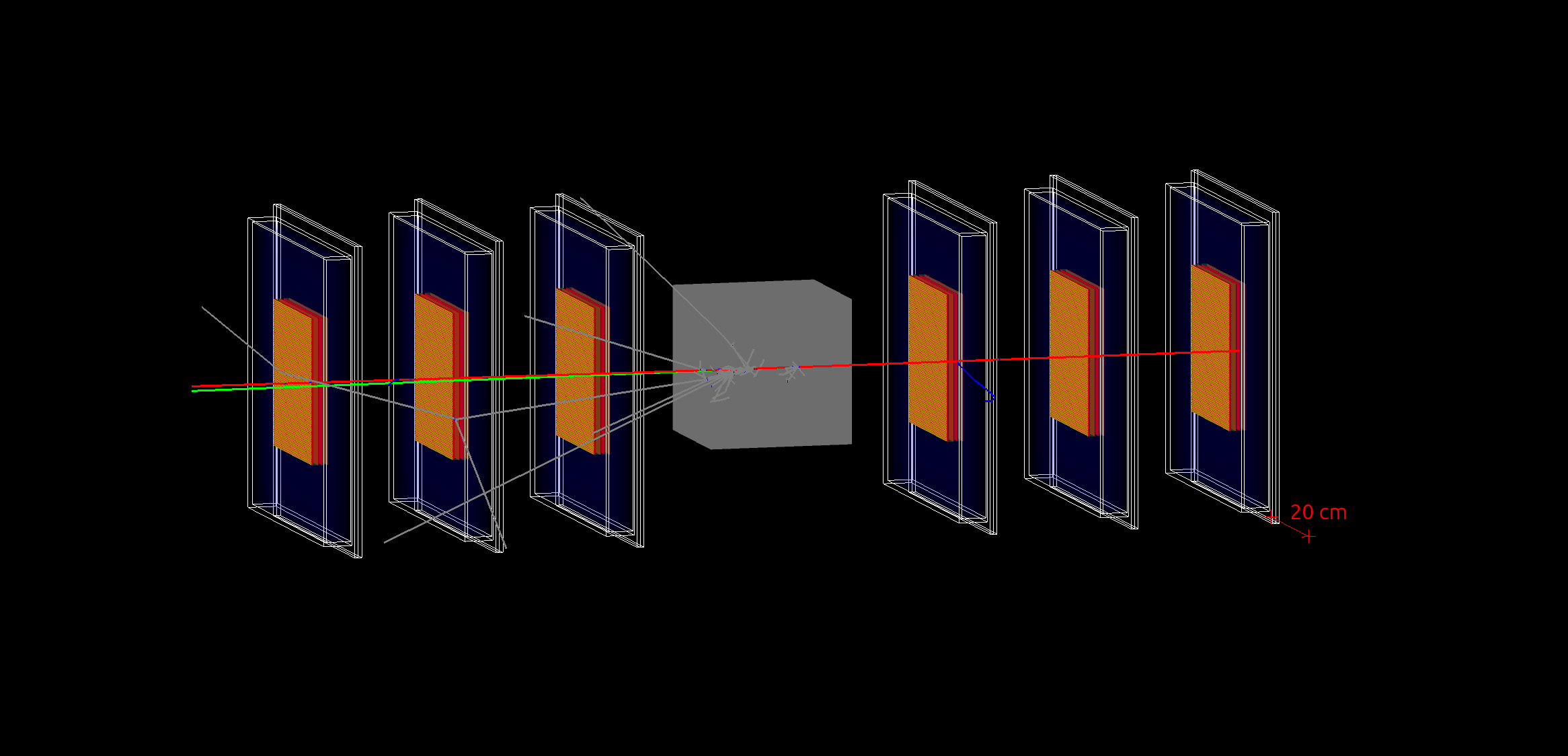}
\caption{\justifying Geometry configuration and ${\color{red}\mu^+}{\color{blue}e^-} \to {\color{red}\mu^+}{\color{green}\mu^-}$ event display of the experiment simulation. Other tracks are shown in gray.}
\label{fig:6RPC}
\end{figure}

As shown in Fig. \ref{fig:6RPC}, the detector system consists of 6 copies of RPCs manufactured by Peking University \footnote{Geometry specification for Peking University RPC detector. \url{https://github.com/PKMuon/PKMUON_2024/blob/clfv/config/rpc.yaml}.}, divided into two groups to measure the momentum directions of the incoming and outgoing particles respectively. The distance between the centers of two adjacent intragroup (intergroup) RPCs is set as $200$ ($500$) $\mathrm{mm}$ (which can be optimised further). Each RPC copy has two sensitive layers with respectively $x$- and $y$-readout systems attached. 2D pixel-like output with time resolution $50\ \mathrm{ns}$ and space resolution $0.5\ \mathrm{mm}$ can be achieved via waveform analysis electronics and is simulated as pixel readout for simplicity. The pixels are sized $0.5\times0.5\ \mathrm{mm^2}$ large and distributed symmetrically about the $x$- and $y$-axis. The range cuts for all kinds of particles are set as $0.7\ \mathrm{mm}$ which is the GEANT4 default \cite{GEANT4:2023application}, e.g., disabling tracking secondary electrons/positrons in the sensitive volumes below kinematic energy $\sim\mathrm{keV}$ for simulation performance acquisition with little precision impairment. Muons and electrons extremely dominate those passing the range cuts in the sensitives. The same detector response to energy deposition of all particles is assumed, which applies naturally to $\mu^\pm$ and $e^\pm$. The energy deposition trigger for each readout is set as $1\ \mathrm{MeV}$ to suppress noises. The threshold is about half the energy loss for a muon to pass one sensitive layer.

Typical materials, e.g., lead and aluminum, with either a relatively high electron density or a relatively low muon-nuclear interaction cross section, are selected as the targets to probe the $\mu^+e^- \to \ell^+\ell^-$ CLFV process. The area towards the incoming muon beam is set as $200\times200\ \mathrm{mm^2}$, the same as the sensitive volumes. The thickness is altered in optimization. The CLFV process can happen outside the target wherever electrons exist, and the probability is proportional to the electron density. In the simulation, the target and the volumes inside the RPCs are the dominant CLFV interaction points, and the latter can be filtered out by applying the requirements for the number of hits on each RPC, as discussed in the next section. With the measurement insensitive to the focusing of the incoming muon beam, the beam is simplified as an ideal single-energy beam perpendicularly towards the detector, distributed evenly in a plane of the same area as the detector at the beginning.

\section{Event Reconstruction and Selection}\label{sec:selection}

\begin{figure}[t]
\subfloat[$\chi^2$]{\label{fig:chi2}\includegraphics[width=.5\columnwidth]{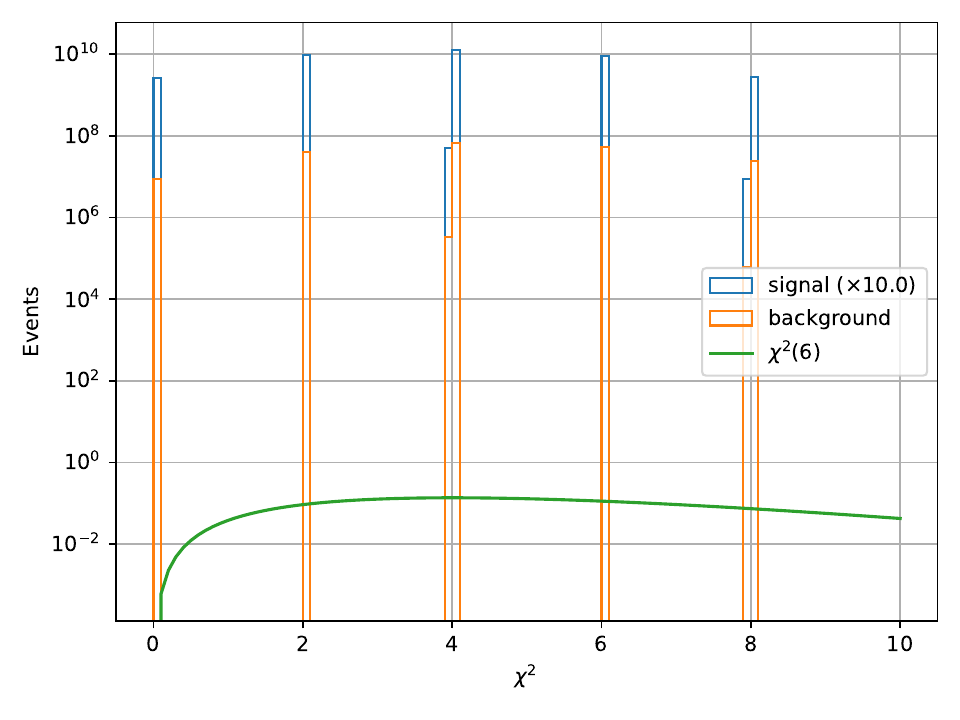}}
\subfloat[$\left<\vec{p}_1, \vec{p}_2\right>$]{\includegraphics[width=.5\columnwidth]{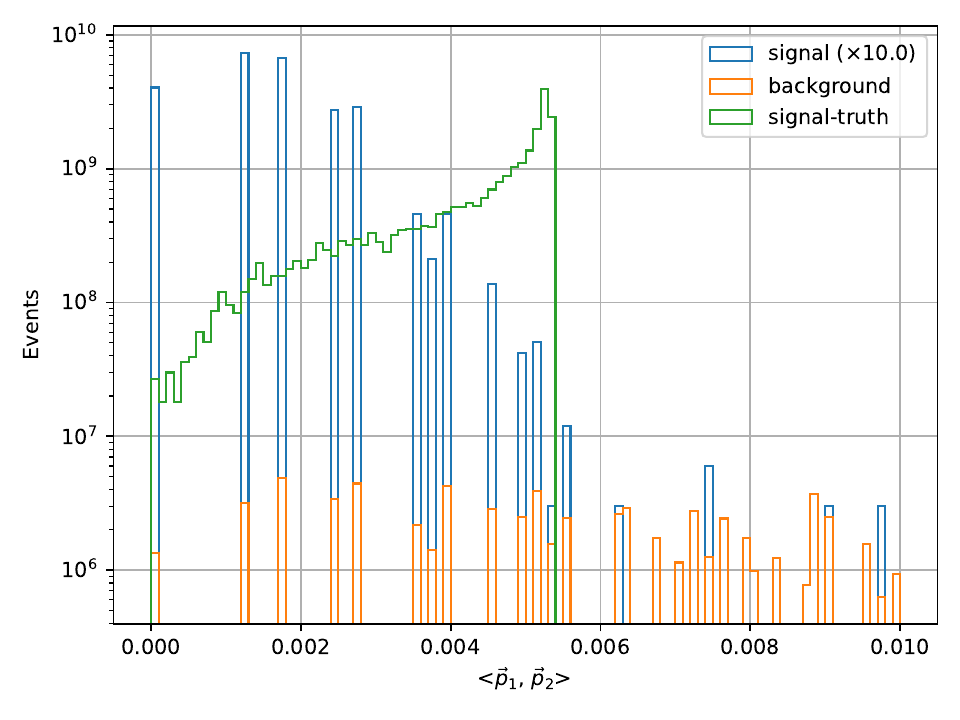}}

\subfloat[$\left<\vec{p}_0, \vec{p}_1\right>$]{\includegraphics[width=.5\columnwidth]{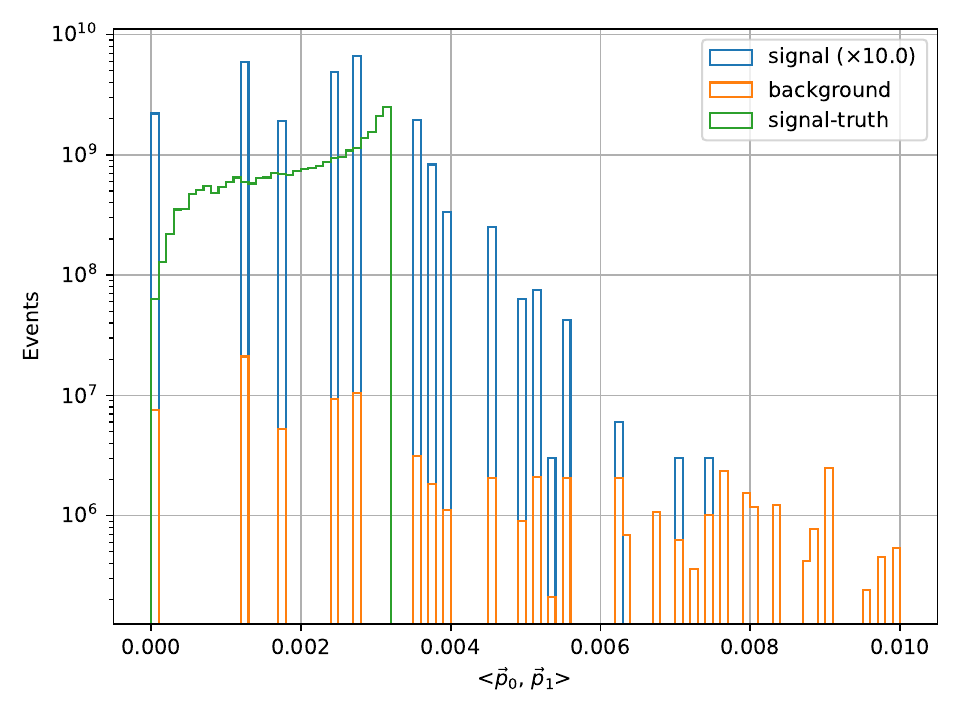}}
\subfloat[$\left<\vec{p}_0, \vec{p}_2\right>$]{\includegraphics[width=.5\columnwidth]{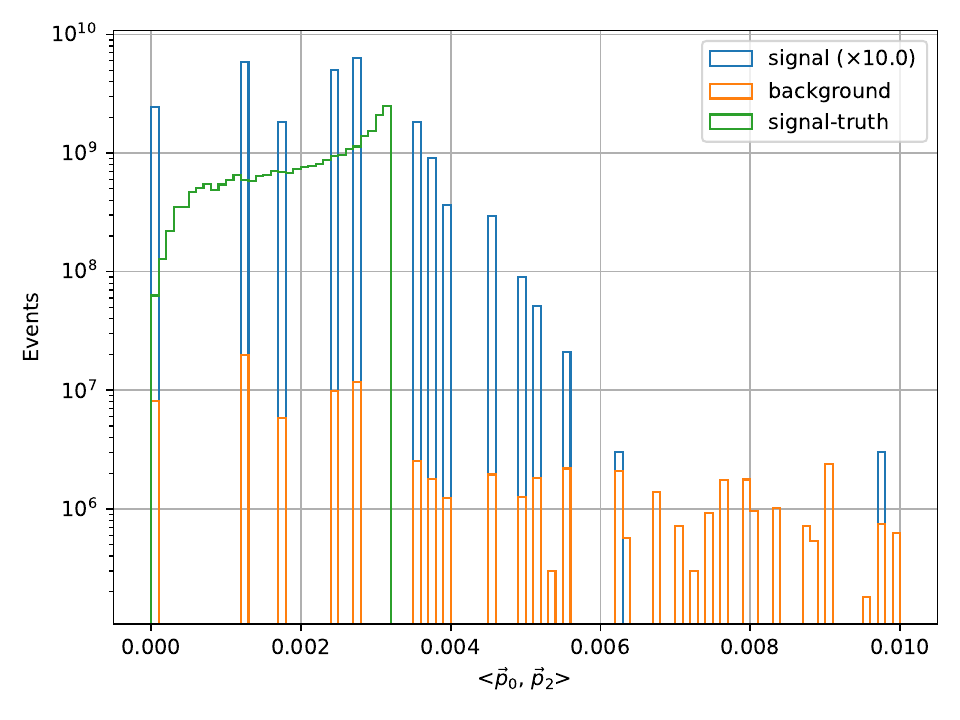}}

\subfloat[$\max\{\left<\vec{p}_0, \vec{p}_1\right>, \left<\vec{p}_0, \vec{p}_2\right>\}$]{\includegraphics[width=.5\columnwidth]{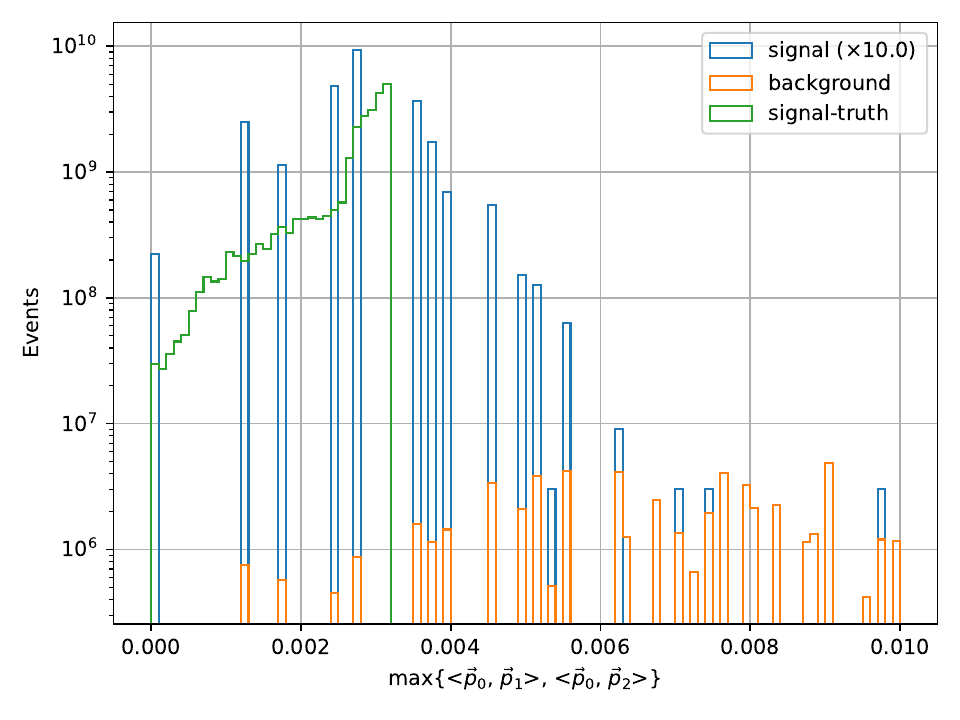}}
\subfloat[$\max_{i \neq j}\{\left<\vec{p}_i, \vec{p}_j\right>\}$]{\includegraphics[width=.5\columnwidth]{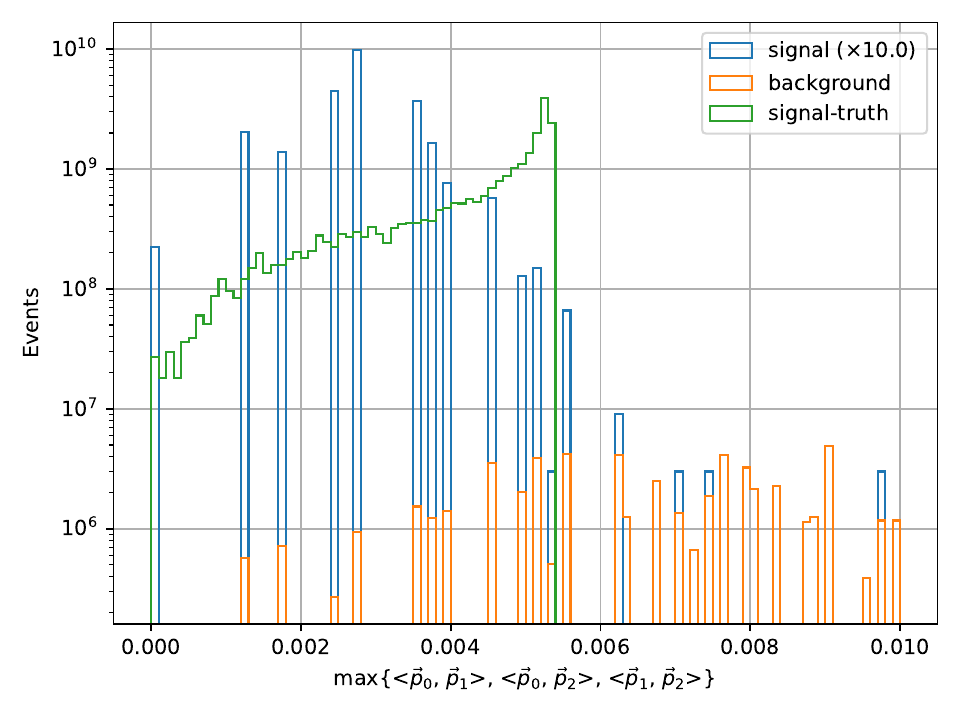}}
\caption{\justifying (a) $\chi^2$ distribution of the events and (b)--(f) angular distributions of the events passing the $\chi^2$ selection, where $E_\mu = 50.2\ \mathrm{GeV}$, $m_{Z^\prime} = 0.25\ \mathrm{GeV}$, the $\mu^+e^- \to \mu^+\mu^-$ process is added with $\lambda_{e\mu}$ scaled to 10, and the target is a $30\ \mathrm{mm}$ thick aluminum board. The yields are normalized to $3 \times 10^{13}$ targeting muons corresponding to a one year's accumulation.}
\label{fig:reco-angs}
\end{figure}

Events with inconsistent hits on the two sensitive layers of any RPC are first dropped as an emulation of the waveform analysis electronics. After that, the 3D coordinates of a hit are reconstructed as the center of two sensitive layers. Then, to precisely reconstruct from the hits 1 incoming and 2 outgoing lepton tracks, for the first (second) group, the number of hits on each RPC is required to be exactly 1 (2) so that the hits on the first and last intragroup RPCs construct 1 (2) tracks. The pairing scheme of the 4 hits in the second group is picked to minimize the sum of the two resulting track lengths, considering that the two tracks diverge from a single interaction point. The next step is to add each hit on each remaining RPC to an individual reconstructed track in the same group, where two combinations are possible for each remaining in the second group. The cluster strategy is to minimize
\begin{equation}
\begin{split}
\chi^2 &= \frac{1}{\sigma_\mathrm t^2}\sum_{t = 1}^{3}\sum_{h = 1}^{3}\left(\hat{\vec r}_{th}\left(\vec r_{t1}, \vec r_{t3}, z_{th}\right) - \vec r_{th}\right)^2
\\ &= \frac{1}{\sigma_\mathrm t^2}\sum_{t = 1}^{3}\left(\hat{\vec r}_{t2}\left(\vec r_{t1}, \vec r_{t3}, z_{t2}\right) - \vec r_{t2}\right)^2
\sim \chi^2(6),
\end{split}
\end{equation}
where $\sigma_\mathrm t = 0.5\ \mathrm{mm} / \sqrt{12}$ is the $x$- and $y$-uncertainty of the detector, $\vec r_{th}$ denotes the $h$-th hit point of the $t$-th track, and $\hat{\vec r}_{th}\left(\vec r_{t1}, \vec r_{t3}, z_{th}\right)$ denotes the point on the spatial line passing points $\vec r_{t1}$ and $\vec r_{t3}$ and of the same $z$ coordinate as point $\vec r_{th}$. Finally, linearly fit the 3 tracks for their more precise directions, recompute $\chi^2$ with the fit tracks in place of the ones determined by the first and last points, and drop the events with $\chi^2 > 6$. An example of the $\chi^2$ distribution in the analysis is shown in Fig. \ref{fig:chi2}. The above procedures reconstruct precisely 3 tracks utilizing all collected trajectory points. Subsequently, the signal event candidates with $\ell = \mu$ of our special interest are further selected based on their distribution in the phase space as described in the following paragraphs, where the momenta of the three tracks are denoted by $\vec p_0$, $\vec p_1$, and $\vec p_2$, as the identities, including the charges of the particles, can not be distinguished by the detector.

For $\ell = \mu$, motivated by the $\theta$ distribution shown in Fig. \ref{fig:theta-alpha} and the fact that two background tracks have little probability to be highly collimated as the outgoing muon pair in a signal event, angular distributions can be taken into account to suppress most backgrounds. Take Fig. \ref{fig:reco-angs} for example. The reconstructed signal and background distributions are discrete because of the limited detector resolution, although the ones in the MC truth are contiguous. The reconstructed signal angles exceed the MC upper limits because of multiple interactions with the detected muons involved in addition to the resolution limitation. The $\left<\vec{p}_0, \vec{p}_1\right>$ and $\left<\vec{p}_0, \vec{p}_2\right>$ distributions are identical as expected by the indistinguishability, and both of them contain comparable signal and background events in the counting range. On the contrary, the $\max\{\left<\vec{p}_0, \vec{p}_1\right>, \left<\vec{p}_0, \vec{p}_2\right>\}$ distribution in the same range contains few background events, as same as the $\left<\vec{p}_1, \vec{p}_2\right>$ distribution. The $\max\{\left<\vec{p}_0, \vec{p}_1\right>, \left<\vec{p}_0, \vec{p}_2\right>\}$ and $\left<\vec{p}_1, \vec{p}_2\right>$ values can be considered collectively for better signal purification in compensation of the limited detector resolution, motivating the selection strategy based on $\max_{i \neq j}\{\left<\vec{p}_i, \vec{p}_j\right>\}$.

\section{Results}\label{sec:results}

\begin{figure}[t]
\subfloat[signal]{\label{fig:reco-contrib-signal}\includegraphics[width=.5\columnwidth]{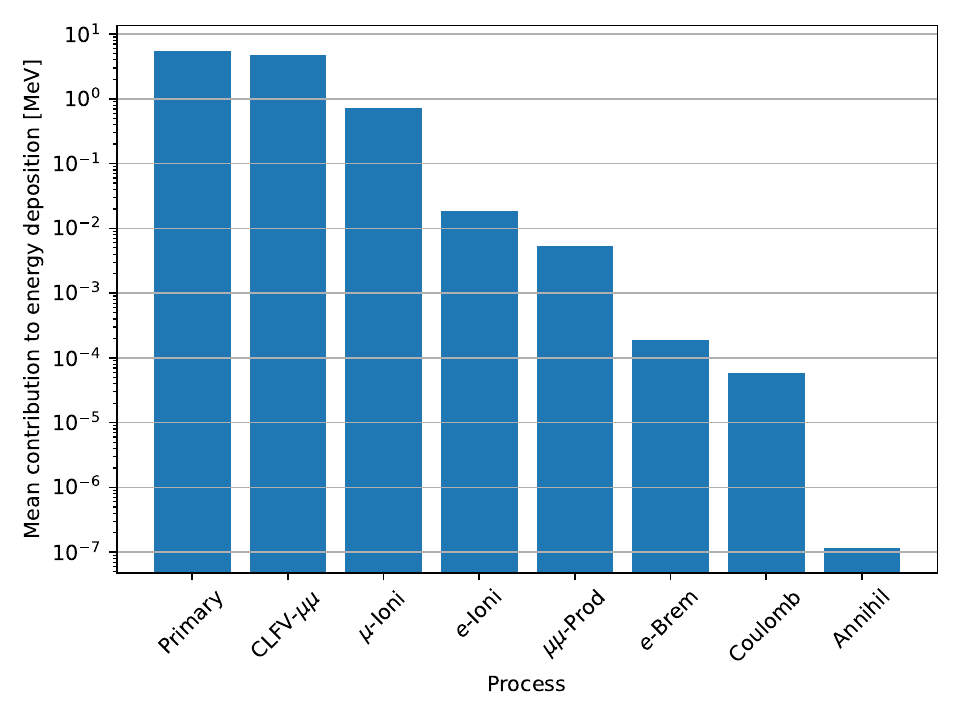}}
\subfloat[background]{\includegraphics[width=.5\columnwidth]{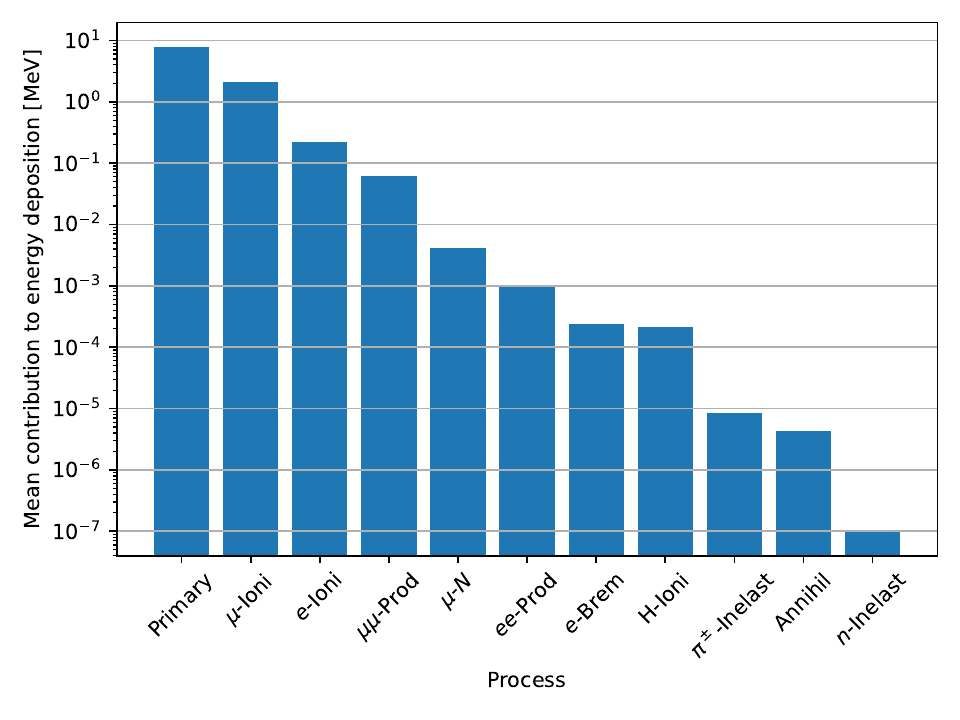}}
\caption{\justifying Contributions to the total energy deposition in all scoring layers of the (a) signal and (b) background events, shown as the average among all events passing the reconstruction and $\chi^2$ selection, where $E_\mu = 50.2\ \mathrm{GeV}$, $m_{Z^\prime} = 0.25\ \mathrm{GeV}$, the $\mu^+e^- \to \mu^+\mu^-$ process is added with $\lambda_{e\mu}$ scaled to 10, and the target is a $30\ \mathrm{mm}$ thick aluminum board. The abbreviations in subfigure (b) are for the primary track ($\mu^+$), muon ionization products, electron/positron ionization products, $\mu^+\mu^-$ pair products, muon-nuclear scattering products, electron-positron pair products, electron/positron bremsstrahlung products, hadron ionization products, and so on.}
\label{fig:reco-contrib}
\end{figure}

Fig. \ref{fig:reco-contrib} shows from the MC truth the mean contributions to energy deposition in the scoring layers of the signal and background events before applying any angular selection. With energy tens of $\mathrm{GeV}$, ionization is the primary form of muon-material interactions, thus contributes most to the scored background energy deposition. Energetic electrons as a part of the ionization products lose energy mainly by secondary ionization and bremsstrahlung. Although there is comparable energy deposition contributed by background processes in the signal events as shown in Fig. \ref{fig:reco-contrib-signal}, about 10\% of the signal events pass the reconstruction and $\chi^2$ selection, depending on and indicating an appropriate readout threshold. After the $\chi^2$ selection, major background contributions except the muon pair products are all electrons/positrons and can be vetoed by an additional scintillator PID system with a negligible false negative rate.

\begin{figure}[t]
\subfloat[$3\ \mathrm{cm}$ Al target]{\label{fig:amm-al}\includegraphics[width=.5\columnwidth]{figure/reco_mup_50.2GeV_Zp_0.25GeV_mumu_x1e+01_al_AMM.pdf}}
\subfloat[$3\ \mathrm{cm}$ Al target, with $e$-veto]{\label{fig:amm-al-ne}\includegraphics[width=.5\columnwidth]{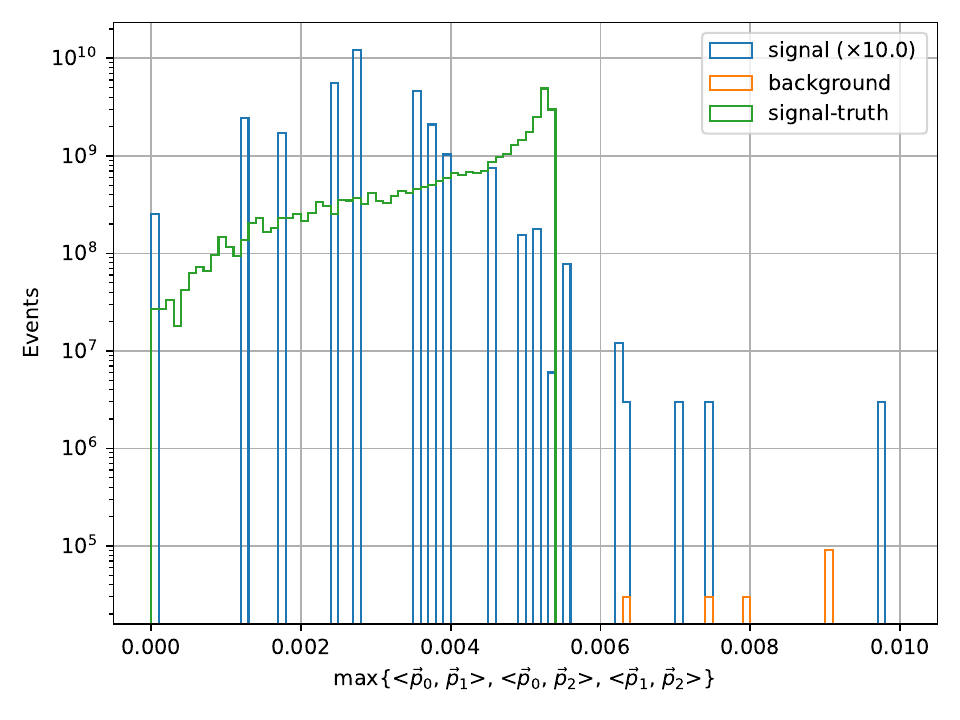}}

\subfloat[$8\ \mathrm{cm}$ Pb target]{\label{fig:amm-pb}\includegraphics[width=.5\columnwidth]{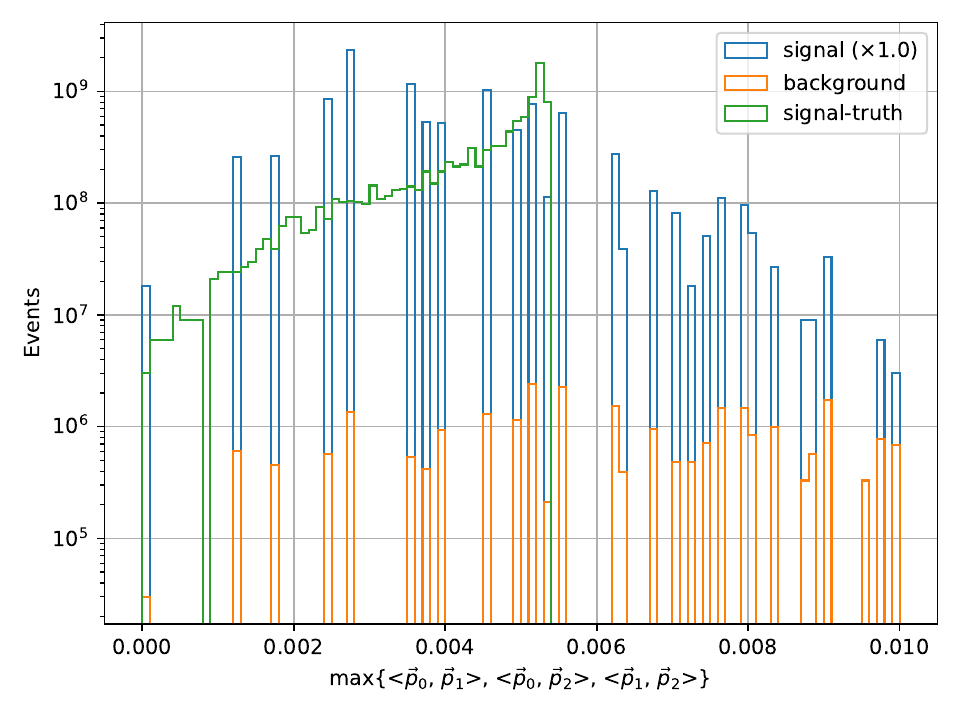}}
\subfloat[$8\ \mathrm{cm}$ Pb target, with $e$-veto]{\label{fig:amm-pb-ne}\includegraphics[width=.5\columnwidth]{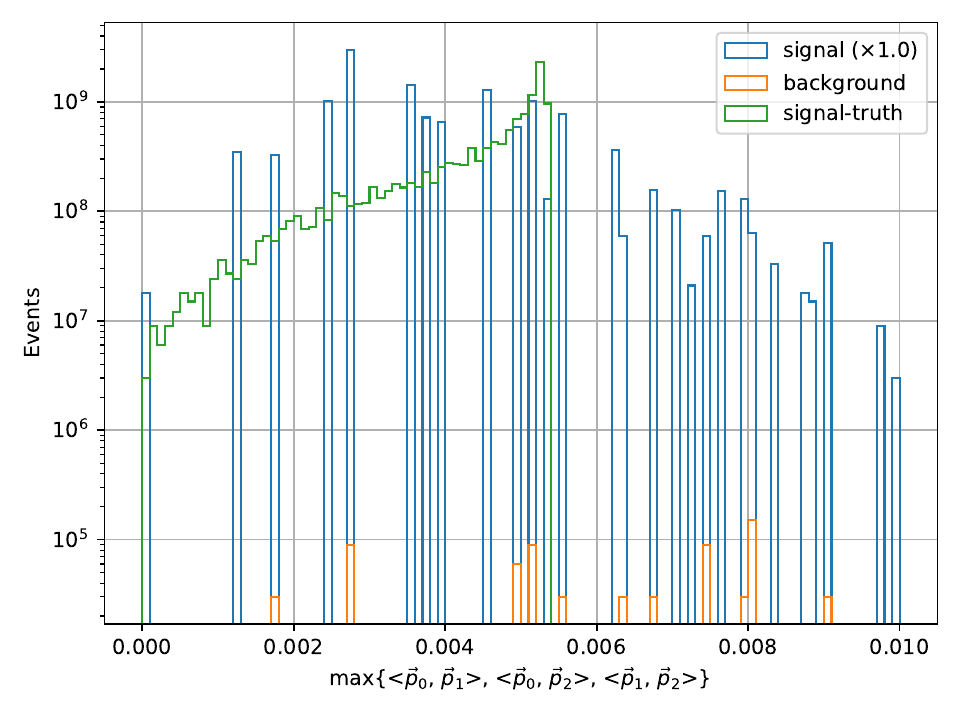}}

\caption{\justifying $\max_{i \neq j}\{\left<\vec{p}_i, \vec{p}_j\right>\}$ distributions of variant target, where $E_\mu = 50.2\ \mathrm{GeV}$, $m_{Z^\prime} = 0.25\ \mathrm{GeV}$, and the $\mu^+e^- \to \mu^+\mu^-$ process is added with $\lambda_{e\mu}$ scaled to 10 (1) for Al (Pb). The yields are normalized to $3 \times 10^{13}$ targeting muons corresponding to a one year's accumulation.}
\label{fig:amm}
\end{figure}

Fig. \ref{fig:amm} shows the $\max_{i \neq j}\{\left<\vec{p}_i, \vec{p}_j\right>\}$ distributions of several selected targets. $10^9$ events are generated for each plot and scaled to the one year's accumulation of $10^6/\mathrm{s} \times 1\,\text{year} = 3\times10^{13}$ incoming muons. Fig. \ref{fig:amm-al-ne} indicates a background-free scenario, while Fig. \ref{fig:amm-pb-ne} shows higher background yields with slightly higher signal yields divided by the global cross section scale factor. The events are selected by requiring $\max_{i \neq j}\{\left<\vec{p}_i, \vec{p}_j\right>\} \leq 0.003$ to suppress out-of-collimation backgrounds. After the selection, as suggested in Ref. \cite{Li:2023lin,Cowan:2010js}, the test statistics $Z$ for 95\% confidence level upper limit is utilized to describe the expected sensitivity of the proposed experiment to the $\mu^+e^- \to \mu^+\mu^-$ CLFV process:
\begin{equation}
Z = 2\left(\mu s + b\log\left(b / (\mu s + b)\right)\right) \sim \chi^2(1),
\end{equation}
where $s$ and $b$ are respectively the post-selection signal and background yields normalized to one year and $\mu$ is the signal strength to be limited by the upper 5\% point of the $\chi^2(1)$ distribution. The $\mu$ result is finally translated to the limit on $(\lambda_{e\mu}\lambda_{\mu\mu})^2$ by multiplying it with the global cross section scale factor applied to the corresponding simulation run, on the basis of $\lambda_{e\mu}\lambda_{\mu\mu} = 1$ set for all runs.

\begin{figure}[t]
\subfloat[$3\ \mathrm{cm}$ Al target, with $e$-veto]{\label{fig:ul-al-ne}\includegraphics[width=.5\columnwidth]{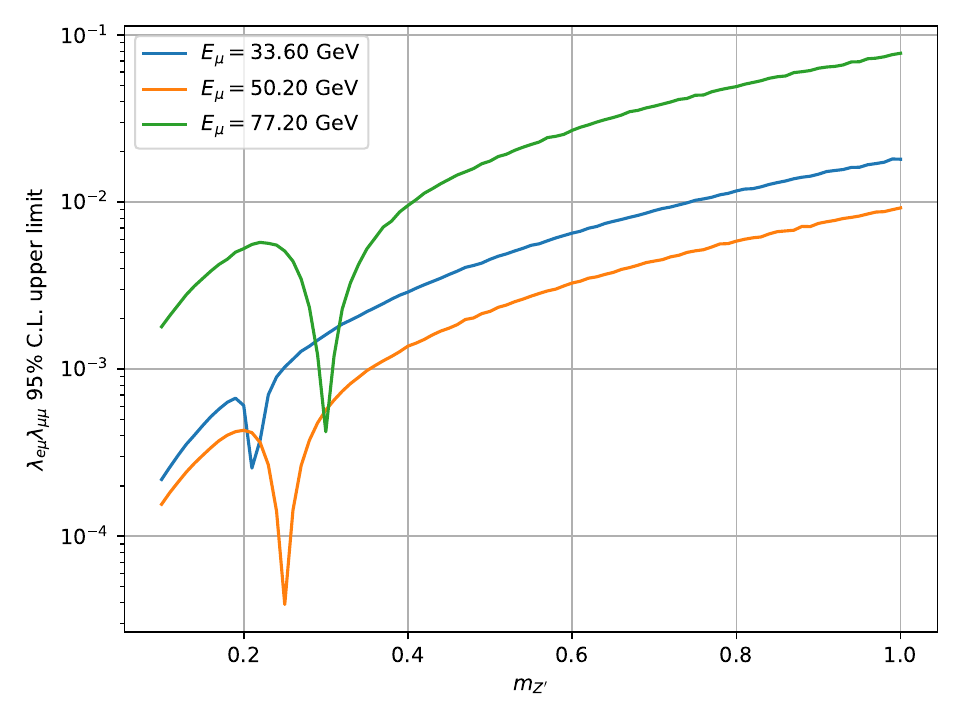}}
\subfloat[$8\ \mathrm{cm}$ Pb target, with $e$-veto]{\label{fig:ul-pb-ne}\includegraphics[width=.5\columnwidth]{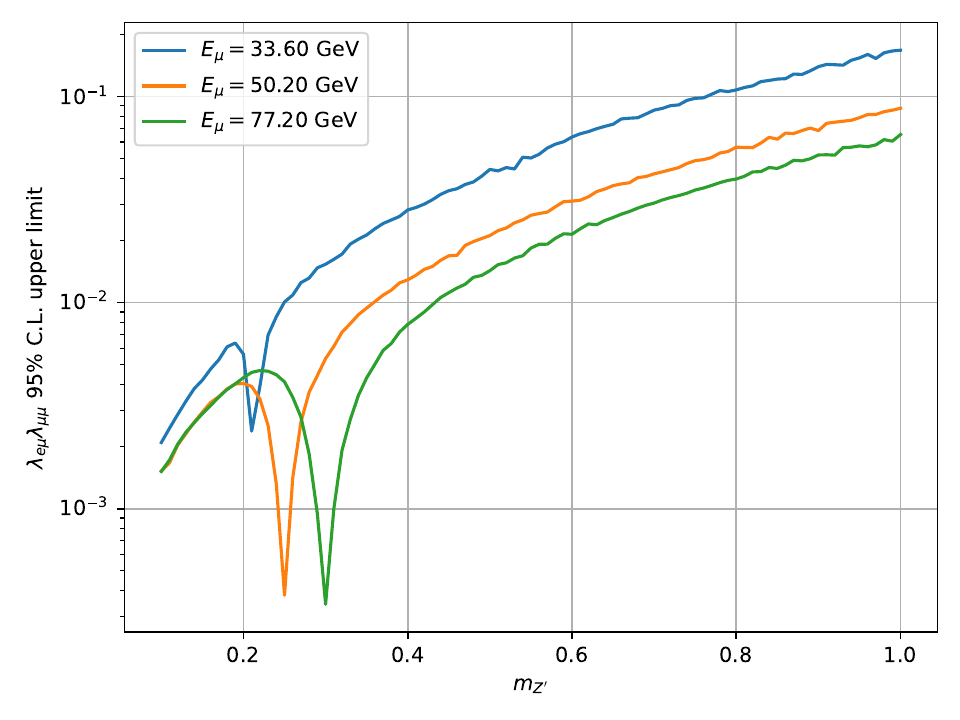}}
\caption{\justifying 95\% C.L. upper limit results of variant target. The yields are normalized to $3 \times 10^{13}$ targeting muons corresponding to a one year's accumulation.}
\label{fig:ul}
\end{figure}

Fig. \ref{fig:ul} shows the expected 95\% confidence level upper limit results of two typical targets with $e$-veto applied. The best result, with $E_\mu = 50.2\ \mathrm{GeV}$ and a $3\ \mathrm{cm}$ thick Al target, is free from background and weaker than Ref. \cite{Ding:2024zaj} by an order of magnitude because of a true positive rate $\sim 10\%$ of the whole event reconstruction and selection procedure. The expected sensitivity for the $8\ \mathrm{cm}$ thick Pb target is weaker with more background contamination, as indicated in Fig. \ref{fig:amm}.

In the above two cases, the background events either vanish after the selection or remain with a high uncertainty caused by the large normalizing scale factor, disabling the fine optimization of the selection. The future enrollment of an even larger amount of computing resource will mitigate the problem and make it possible to optimize for a better sensitivity. And in addition to the current reconstruction and selection requirements, there are more options available, e.g., the reconstructed CLFV candidate interaction position and the invariant mass of the two outgoing leptons. They are also typical signal characteristics and help to further purify the signal.

\section{Summary and Outlook}\label{sec:summary}

Detailed simulation studies are performed for probing the $Z^\prime$-mediated CLFV process in a proposed muon on-target experiment based on the Peking University Muon Experiment facilities. A special event generation algorithm cooperated with the GEANT4 simulation framework is proposed and implemented to enable efficient simulation in a sandwich-like muon detecting system and is generalizable to many other physics processes of great interest such as muon-electron scattering processes to the final states with invisible, $\ell$ + invisible, or $\ell\ell$ (+ invisible). Results of the $\mu\mu$ final state are shown that the expected 95\% confidence level upper limit on the coupling coefficients $\lambda_{e\mu}\lambda_{\mu\mu}$ is up to $10^{-5}$ with $Z^\prime$ mass $0.25\ \mathrm{GeV}$ for a one year's run. This, in addition to previous limits (see discussions in Ref.~\cite{Ding:2024zaj}) from muon decay or conversion experiments, can provide necessary complementary information to probe certain CLFV couplings exclusively. Notice one can also start as the first step from the cosmic muon targeting events where the detecting system can be built upon the current running one in the PKMuon Experiment and the simulation framework developed for this work can be easily ported to the cosmic source.

\begin{acknowledgments}
This work is supported in part by the National Natural Science Foundation of China under Grants No. 12150005, No. 12325504, and No. 12075004.
\end{acknowledgments}

\appendix

\section{Kinematic Relations for Outgoing Leptons in the COM and Lab Frame}\label{sec:kinematic}

For $\mu^+e^- \to \ell^+\ell^-$ process, in the lab frame, as shown in Fig. \ref{fig:lab-frame}, given the incoming $\mu^+$ towards the positive direction of the $z$ axis with total energy $E_\mu$ and the target $e^-$ approximately at rest, the energy-momentum of the system can be calculated as
$
E = E_\mu + m_e,\
p = \sqrt{E_\mu^2 - m_\mu^2},
$
leading to the COM energy
$
E^\prime = \sqrt{E^2 - p^2} = \sqrt{2E_\mu m_e + m_e^2 + m_\mu^2}.
$
This gives the Lorentz boost factors of the COM frame in the lab frame along the z axis as
$
\gamma = E / E^\prime,\
\beta = p / E,
$
determining the Lorentz transformation from the COM frame to the lab frame as
\begin{equation}\label{eq:lorentz}
p_\mathrm t = p^\prime_\mathrm t,\
p_z = \gamma\left(p^\prime_z + \beta E^\prime\right),\
E = \gamma\left(E^\prime + \beta p^\prime_z\right).
\end{equation}

In the COM frame, as shown in Fig. \ref{fig:com-frame}, the two outgoing leptons always have the momenta with exactly the same magnitude $p^\prime = \sqrt{{E^\prime}^2 / 4 - m_\ell^2}$ and the opposite directions, characterised by $\alpha$, the polar angle of $\ell^+$. Transforming $\left(\vec p_+^{\,\prime}, E^\prime/2\right)$ to the lab frame using Eq. \eqref{eq:lorentz}, we get
$
p_{\mathrm t+} = p^\prime\sin\alpha,\
p_{\mathrm z+} = \gamma\left(p^\prime\cos\alpha + \beta E^\prime/2\right).
$

\begin{figure}[t]
\centering
\subfloat[\mbox{$\mu^+e^- \to e^+e^-$    }]{\includegraphics[width=.5\columnwidth]{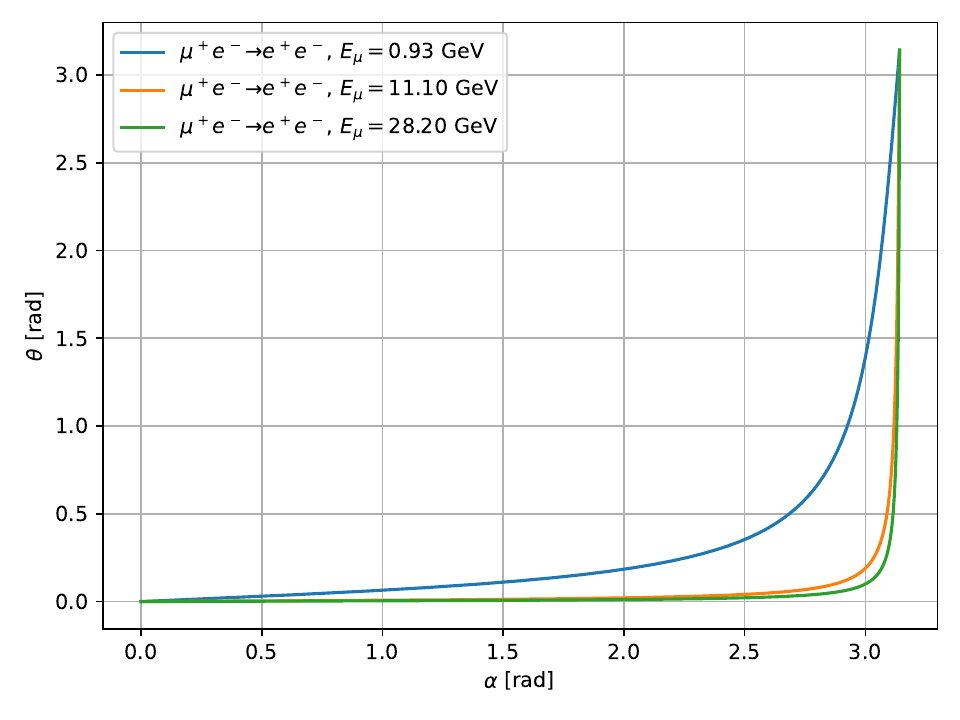}}
\subfloat[\mbox{$\mu^+e^- \to \mu^+\mu^-$}]{\includegraphics[width=.5\columnwidth]{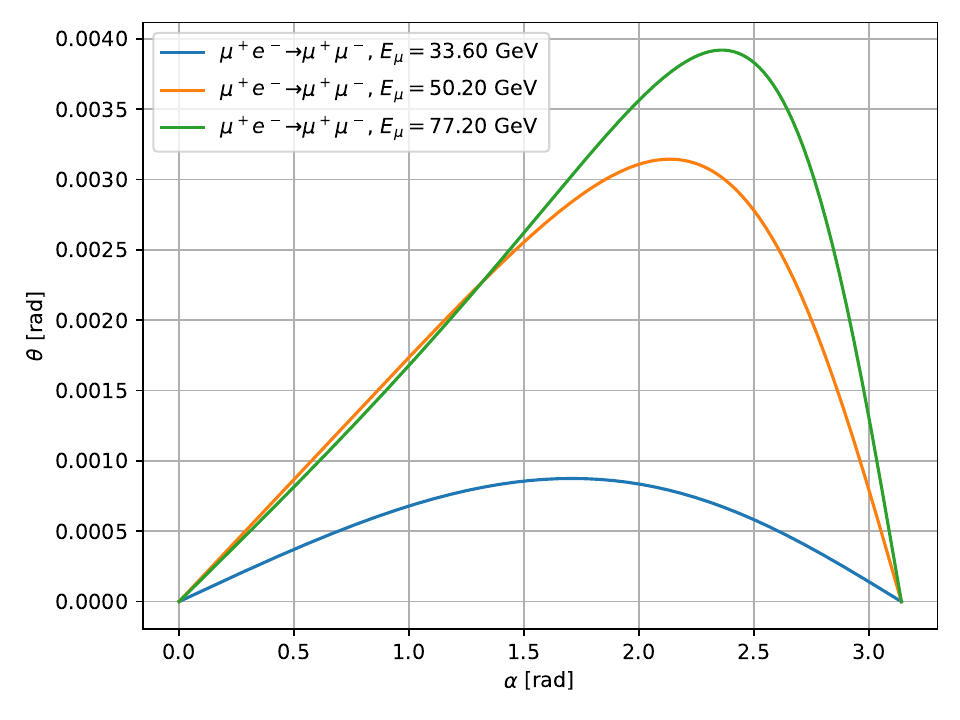}}
\caption{\justifying Typical $\theta$-$\alpha$ relationships in $\mu^+e^- \to \ell^+\ell^-$ processes, obtained by Eq. \eqref{eq:alpha}.}
\label{fig:theta-alpha}
\end{figure}

On the one hand, given $\theta$, the polar angle of $\ell^+$ in the lab frame, we have
\begin{equation}\label{eq:alpha}
\tan\theta = \frac{p_{\mathrm t+}}{p_{\mathrm z+}} = \frac{p^\prime\sin\alpha}{\gamma\left(p^\prime\cos\alpha + \beta E^\prime/2\right)},
\end{equation}
where the only unknown variable $\alpha$ can be solved, giving
$$
\cos\alpha = \frac{-A^2B \pm \sqrt{\left(A^2B\right)^2 - \left(1 + A^2\right)\left(A^2B^2 - 1\right)}}{1 + A^2},
$$
where $A = \gamma\tan\theta$, $B = \beta E^\prime / \left(2p^\prime\right)$ are both known constants. Generally, there can be one or two viable $\alpha$ solutions, as shown in Fig. \ref{fig:theta-alpha}. The final state thus remains nondeterministic because of the non-unique $\alpha$ solution.

On the other hand, given $\vec p_+$, the momentum of the outgoing $\ell^+$, instead of only its polar angle $\theta$ in the lab frame, the polar angle $\alpha$ in the COM frame can be calculated as
\begin{equation}
\tan\alpha = \frac{p_{\mathrm t+}^\prime}{p_{\mathrm z+}^\prime}
= \frac{p_{\mathrm t+}}{\gamma\left(p_{\mathrm z+} - \beta\sqrt{p_+^2 + m_\ell^2}\right)},
\end{equation}
where the inverse Lorentz transformation against Eq. \eqref{eq:lorentz} is applied. The final state thus becomes deterministic after e.g., giving the azimuthal angle of the outgoing $\mu^+$, which is of little significance in the system illustrated in Fig. \ref{fig:frames} with rotational symmetry.

We conclude from the above that $\alpha$, the polar angle of the outgoing $\ell^+$ in the COM frame, whose distribution can be obtained from $\vec p_+$ in the lab frame, uniquely describes the $\mu^+e^- \to \ell^+\ell^-$ process, without considering the trivial degree of freedom resulting from the azimuthal angle, while generally the ambiguity cannot be avoided by using only $\theta$, the same angle in the lab frame.


\bibliographystyle{apsrev4-2}
\bibliography{apssamp}

\end{document}